# Terahertz Optics Driven Phase Transition in Two-Dimensional Multiferroics

Jian Zhou[1,*], Shunhong Zhang[2,†]

[1] *Center for Advancing Materials Performance from the Nanoscale, State Key Laboratory for Mechanical Behavior of Materials, Xi'an Jiaotong University, Xi'an, 710049, China*

[2] *International Center for Quantum Design of Functional Materials (ICQD), Hefei National Laboratory for Physical Sciences at Microscale, and CAS Center For Excellence in Quantum Information and Quantum Physics, University of Science and Technology of China, Hefei, Anhui 230026, China*

Abstract

Displacive martensitic phase transition is potentially promising in semiconductor based data storage applications with fast switching speed. In addition to traditional phase transition materials, the recently discovered two-dimensional ferroic materials are receiving lots of attention owing to their fast ferroic switching dynamics, which could tremendously boost data storage density and enhance read/write speed. In this study, we propose that a terahertz laser with an intermediate intensity and selected frequency can trigger ferroic order switching in two-dimensional multiferroics, which is a damage-free noncontacting approach. Through first-principles calculations, we theoretically and computationally investigate optically induced electronic, phononic, and mechanical responses of two experimentally fabricated multiferroic (with both ferroelastic and ferroelectric) materials, $\beta$-GeSe and $\alpha$-SnTe monolayer. We show that the relative stability of different orientation variants can be effectively manipulated via the polarization direction of the terahertz laser, which is selectively and strongly coupled with the transverse optical phonon modes. The transition from one orientation variant to another can be barrierless, indicating ultrafast transition kinetics and the conventional nucleation-growth phase transition process can be avoidable.

**Introduction.**

Optical control of material geometries is receiving rapidly growing attentions in very recent years. For example, some ferroelectric/multiferroic perovskites, such as $BaTiO_3$[1,2], $BiFeO_3$[3], and $SrTiO_3$[4,5] would experience their ferroic order switch under laser pulse illumination (with the optical frequency well-below their electronic bandgaps). It offers a remarkable and promising scheme to manipulate the structure of materials, avoiding mechanical or electrochemical contacts with these samples, which might slow down the effect and introduce unwanted impurities or disorders. Thus, this fast-paced noncontacting opto-mechanical strategy is less susceptible to lattice damage and provides an ultrafast manipulation of materials on picosecond time scales and sub-micrometer length scales[6].

According to optical response theory, there are four regimes of optical frequency, namely, low frequency regime where optical frequency $\omega < \omega_0 - \frac{1}{\tau}$ ($\omega_0$ is energy difference between two states and $\tau$ is lifetime), absorption frequency regime where $\omega_0 - \frac{1}{\tau} < \omega < \omega_0 + \frac{1}{\tau}$, reflection regime where $\omega_0 + \frac{1}{\tau} < \omega < \omega_p$ (where $\omega_p$ is plasmon frequency), and transparent regime where $\omega > \omega_p$. The second frequency regime may introduce significant heat, the light is highly reflected in the third frequency regime, and the optical response in transparent regime is usually small. Hence, low optical energy light would be intriguing to control the material behaviors practically[7]. Low optical energy light can be further classified into two categories: near or mid-infrared optics with its photon energy above phonon but below electronic bandgap of semiconducting materials [such as experiments on $BaTiO_3$, Ref. (1,2)], and far infrared optics with its (terahertz) frequency strongly and directly coupled with phonons.

Atomically thin two-dimensional (2D) materials, with their extremely large surface area-to-volume ratio, are more optically addressable and accessible. Therefore, noncontacting optically driven ferroic order transition in 2D ferroic materials will be promising with their easy and damage-free manipulation, large information storage density, and ultrafast kinetics. Previous theoretical studies mainly use parameterized model including phonon-phonon nonlinear interactions[8,9]. In this work, we theoretically and computationally evaluate the interactions between light and phonons, as well as light and electrons in 2D time-reversal invariant multiferroic (with both ferroelectric and ferroelastic orders) materials, from a first-principles approach. We choose two experimentally fabricated systems, i.e., monolayers $\beta$-GeSe[10] and $\alpha$-SnTe[11], to illustrate our theory. We predict that linearly polarized terahertz laser (LPTL) pulses with intermediate intensity (around 1–2 MV/cm) can trigger ferroic order switch in these systems. A contact-free direction-dependent vibrational electron energy loss spectroscopy (EELS) is also theoretically calculated, which can be used to detect and measure the structural signature in a high resolution. In addition, we calculate the second harmonic generation (SHG) effects, which also couple to ferroic orders. These strategies could provide powerful and non-invasive tools to characterize ferroicity that is indistinguishable by the traditional diffraction methods.

We analyze the light-matter interaction thermodynamically. When LPTL is illuminated onto a semiconductor (with optical bandgap larger than ~40 meV), the electron-hole pair formation is eliminated owing to small photon energy. Therefore, only optical electromagnetic field effects need to be considered. Here we are interested in the time-reversal invariant systems, in which the magnetic field interaction is very weak and can be omitted. Hence, we focus on the alternating electric field component ($\vec{\mathcal{E}} = \vec{E}_{\max} e^{i\omega t}$, $\omega \sim$ THz) effect, here $E_{\max}$ is the maximum value of the electric field. The LPTL would accelerate both electronic and phononic subsystems in the material, and its work done per volume can be evaluated by $du = \text{Re}(\vec{\mathcal{E}} \cdot d\vec{\mathcal{P}}^*)$, where $\vec{\mathcal{P}}$ is time dependent polarization density. Here, closed boundary condition[12,13] is applied, since electric polarization is in the *xy*-plane. Using Legendre transformation, the Gibbs free energy (GFE) density change is then $d\mathcal{G} = -\text{Re}\langle\vec{\mathcal{P}}^* \cdot d\vec{\mathcal{E}}\rangle$, and $\langle \cdot \rangle$ indicates time average. Note that $\vec{\mathcal{P}} = \vec{P}_s + \varepsilon_0 \text{Re}\overleftrightarrow{\chi}(\omega) \cdot \vec{\mathcal{E}}$, where $\vec{P}_s$ is spontaneous electric polarization, $\varepsilon_0$ is vacuum permittivity, and $\overleftrightarrow{\chi}(\omega)$ is optical susceptibility tensor at the frequency $\omega$, containing electronic and phononic contributions ($\overleftrightarrow{\chi} = \overleftrightarrow{\chi}^{\text{el}} + \overleftrightarrow{\chi}^{\text{ph}}$). Under LPTL illumination (along the *i*-direction), note that the light frequency is on the THz order, which could be lower than the phonon Debye frequency of the system. Hence, if $E_{max}$ is large enough (much stronger than the coercive field $E_c$ to reverse polarization), the spontaneous polarization may follow the electric field and switch back and forth between $\vec{P}_s$ and $-\vec{P}_s$. When the $E_{max}$ is much stronger than $E_c$ (corresponds to the situation in our current discussion), we can show that it contributes to the time-averaged GFE density in the form of (see Supplementary Note 3 for details)

$$g_0 = -\frac{E_{max} P_s \cos\theta}{2}, \quad (1)$$

where $\theta \in [0, \frac{\pi}{2}]$ is the angle between the LPTL polarization direction and the spontaneous polarization direction. This is a first order interaction that measures between light and polarization. If the light frequency is highly above the phonon frequency range (e.g. several tens THz and above), then the ion position change cannot follow the light field (off-resonant). Then this term would become zero as the $\vec{\mathcal{P}}_s$ is time-independent. We also include the second order interactions by incorporating optical susceptibility. This can be considered by the averaged GFE density change through integrating time and electric field in $dg = -\varepsilon_0 \vec{\mathcal{E}}^*(t) \cdot \text{Re}[\overleftrightarrow{\chi}^{el}(\omega) + \overleftrightarrow{\chi}^{ph}(\omega)] \cdot d\vec{\mathcal{E}}(t)$, and the total GFE density contributed from optical linear responses can be written as

$$g_1(\omega) = -\frac{1}{4}\varepsilon_0 \vec{E}_{max} \cdot \text{Re}[\overleftrightarrow{\chi}^{el}(\omega) + \overleftrightarrow{\chi}^{ph}(\omega)] \cdot \vec{E}_{max}$$
$$= -\frac{1}{4}\varepsilon_0 \text{Re}[\chi^{el}_{ii}(\omega) + \chi^{ph}_{ii}(\omega)] E^2_{max,i} \quad (2)$$

Note that the integration of $g_1$ over electric field space gives one $\frac{1}{2}$ factor, and the time average of sinusoidal electric field gives another $\frac{1}{2}$. Hence, there is a $\frac{1}{4}$ coefficient in Eq. (2). The total GFE density change under light can then be estimated by $g = g_0 + g_1$. In a ferroic material with multiple ferroic orders, different orders are subject to a simple structural operator (such as a proper or improper rotation $\hat{O}$), and hence are energetically degenerate. Certain directional LPTL could break such degeneracy (Eqs. 1 and 2), and one can expect optically induced ferroic order switch from a metastable order with higher GFE to a stable order with lower GFE. This mechanism is illustrated in Fig. 1.

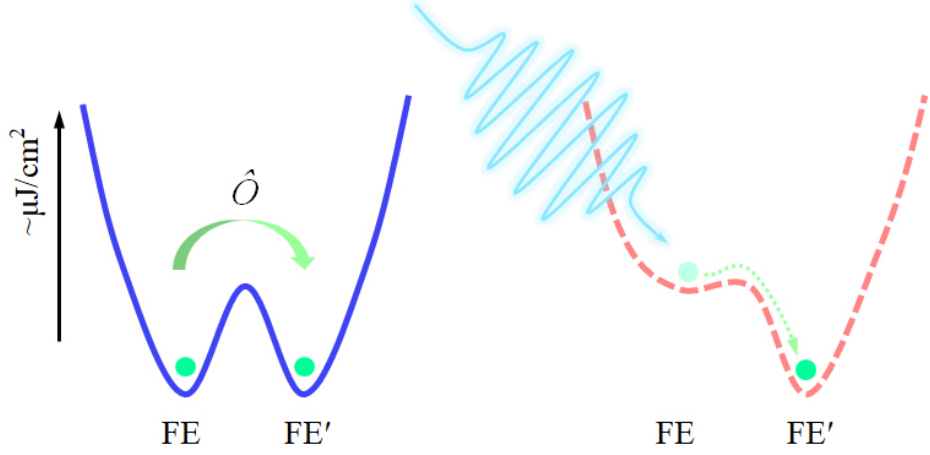

**Fig. 1 LPTL inducing ferroic order switch.** (Left panel) Different ferroic orders (FE and FE′) are subject to structural operator $\hat{\mathcal{O}}$, which are degenerate in energy. (Right panel) Under LPTL, FE′ has lower GFE, corresponding to a FE→FE′ transition.

The optical susceptibility can be evaluated via *ab initio* density functional theory (DFT) calculations (see Computational Methods for details[14-21]). We use the Vienna *ab initio* simulation package[15] to calculate the electron behaviors self-consistently, and compute forces on ions to evaluate phonon dispersions using density functional perturbation theory (DFPT), as implemented in the Phonopy package[21].

According to the linear response theory with random phase approximation (RPA), the electronic part of susceptibility $\overleftrightarrow{\chi}^{\mathrm{el}}(\mathbf{q}=\mathbf{0},\omega)$ is the Fourier transformation of real space density response function $\overleftrightarrow{\chi}^{\mathrm{el}}(\mathbf{r},\mathbf{r}',\omega)$, which is solved from a Dyson equation

$$\overleftrightarrow{\chi}^{\mathrm{el}}(\mathbf{r},\mathbf{r}',\omega) = \overleftrightarrow{\chi}^{\mathrm{el},(0)}(\mathbf{r},\mathbf{r}',\omega) + \int d\mathbf{r}_1 d\mathbf{r}_2 \overleftrightarrow{\chi}^{\mathrm{el},(0)}(\mathbf{r},\mathbf{r}_1,\omega)\frac{1}{|\mathbf{r}_1-\mathbf{r}_2|}\overleftrightarrow{\chi}^{\mathrm{el},(0)}(\mathbf{r}_2,\mathbf{r}',\omega), \quad (3)$$

where $\overleftrightarrow{\chi}^{\mathrm{el},(0)}$ is bare density response function (independent particle) contributed by electron transitions

$$\overleftrightarrow{\chi}^{\mathrm{el},(0)}(\mathbf{r},\mathbf{r}',\omega) = \sum_{m\neq n}(f_n - f_m)\frac{\varphi_m^*(\mathbf{r})\varphi_n^*(\mathbf{r}')\varphi_n(\mathbf{r})\varphi_m(\mathbf{r}')}{\hbar\omega - \epsilon_m + \epsilon_n + i\xi}. \quad (4)$$

Here, $f_i$, $\epsilon_i$, $\varphi_i$ are the Fermi-Dirac distribution, energy, and wavefunction of the *i*-th level, and $\xi$ is the Lorentzian phenomenological damping parameter (taken to be 0.025 eV in our calculations), representing disorder, finite temperature, and impurity effects.

When the frequency of LPTL lies in the range of phonon frequency (a few THz),

vibrational phononic contribution to optical susceptibility needs to be included. According to lattice dynamics theory[22], the LPTL is coupled with infrared-active transverse optical (TO) mode of phonons at the Γ-point of the Brillouin zone. The phononic contributions to susceptibility is calculated according to[23]

$$\chi_{\alpha\beta}^{\text{ph}}(\omega) = \frac{1}{\Omega} \sum_m \frac{\mathcal{Z}_{m,\alpha}^* \mathcal{Z}_{m,\beta}^*}{\omega_m^2 - (\omega + i\Gamma)^2}, \qquad (5)$$

where $\Omega$ is the unit cell volume, $\alpha$ and $\beta$ (= 1, 2, 3) are Cartesian-coordinates indices, $m$ is TO phonon mode index. The Born effective charge of each phonon mode is evaluated as $\mathcal{Z}_{m,\alpha}^* = \sum_{\kappa,\alpha'} Z_{\kappa,\alpha\alpha'}^* u_{m,\kappa\alpha'}$, where $\kappa$ is ion index, $u$ is mass normalized displacement, and $Z_{\kappa,\alpha\alpha'}^*$ is Born effective charge component on each ion. The numerator accounts for the vibrational mode oscillator strengths. Finally, $\Gamma$ is linewidth parameter accounting for the finite lifetime ($\tau = \Gamma^{-1}$) of the vibration. For simplicity, we choose a universal value of $\Gamma$ = 4 cm$^{-1}$, which is comparable with the linewidth determined by phonon-phonon couplings[24,25].

**Results.**

**Monolayer *β*-GeSe.** We now apply these analyses to two experimentally fabricated 2D multiferroic materials. Figure 2(a) shows the atomic structure of *β*-GeSe monolayer, with relaxed lattice constants to be *a* = 3.59 Å and *b* = 5.73 Å. This structure belongs to *Pmn*2$_1$ space group (no. 31) without centrosymmetry, consistent with previous works[26,27]. One clearly observes that it looks like a distorted honeycomb lattice (such as h-BN, compressed along the *y*-direction). Actually, the honeycomb structure is serving as a parental phase (Supplementary Note 4), and the *β*-GeSe shown in Fig. 2(a) is one of its ferroelastic orientation variants (denoted as FE0). The (*a*×*b*) rectangle unit cell in *β*-GeSe corresponds to the (1×√3) supercell of the high symmetric parental structure, from which a spontaneous structural transformation occurs, forming different orientation variants. Similar as the 1T and 1T′ phases in transition-metal dichalcogenide monolayers[28], there are three symmetrically equivalent ferroelastic *β*-GeSe (FE0, FE1, and FE2, subjecting to 120°-rotations). The 2D transformation strain tensor of these orientation variants are $\vec{\eta}_0 = \begin{pmatrix} 0.042 & 0 \\ 0 & -0.040 \end{pmatrix}$, $\vec{\eta}_1 = \begin{pmatrix} -0.016 & -0.041 \\ -0.041 & 0.021 \end{pmatrix}$, and $\overleftrightarrow{\eta}_2 = \begin{pmatrix} -0.016 & 0.041 \\ 0.041 & 0.021 \end{pmatrix}$. Therefore, the switch strain from one orientation variant to another is small, which could occur under intermediate energy input in experiments (Supplementary Note 4). Since they are symmetrically equivalent, we will only calculate physical properties of the FE0. One could perform rotational operation for the other two orientation variants.

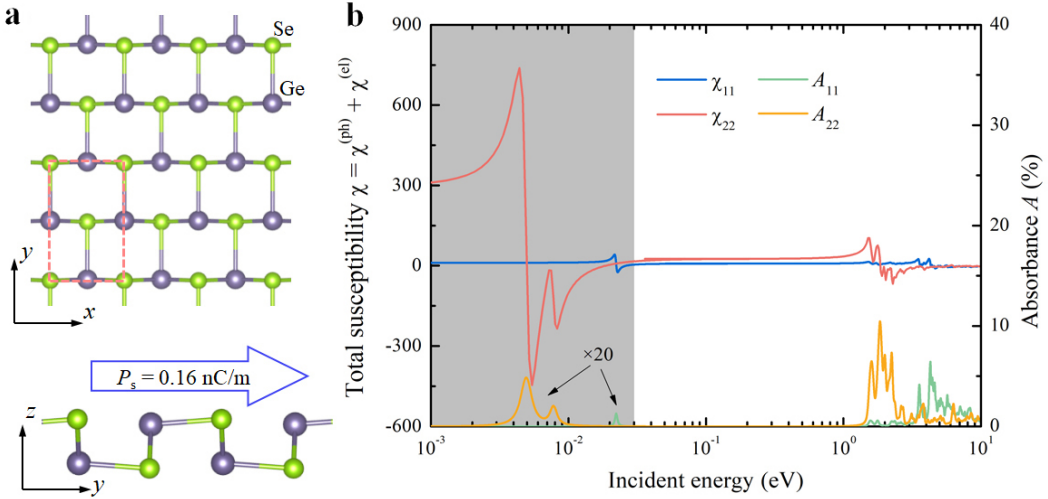

**Fig. 2 Atomic and THz optical responses of monolayer *β*-GeSe.** (a) Geometric structure of monolayer *β*-GeSe, with dashed rectangle indicating simulation unit cell. (b) Calculated real part of optical susceptibility and absorbance function with respect to incident energy. The gray shaded region indicates phononic range (< 8 THz), above which the responses are mainly contributed from electronic subsystem. The absorbance in the phononic range is enlarged twenty times for clarity reasons. The subscripts "1" and "2" denote the Cartesian coordinates "*x*" and "*y*", respectively.

In addition to ferroelastic order, the lack of centrosymmetry indicates a spontaneous electric polarization $\vec{P}_s$. Consistent with previous works[27], we compute its value to be 0.16 nC/m (see Supplementary Note 2 for details), along the armchair (*y*-) direction (Fig. 2a). This polarization is switchable under intermediate in-plane static electric field. Thus, it also possesses a ferroelectric order, making it a 2D time-reversal invariant multiferroic material.

Next, we calculate the electronic and phonon band dispersions (Supplementary Note 1), based on which we can compute the optical susceptibility of FE0-*β*-GeSe, according to Eqs. (3) and (4). Note that all calculations are performed in a 3D periodic supercell with a large vacuum space separating different images. In order to eliminate the vacuum effects and obtain 2D values, we adopt a widely-used scaling method for the real part susceptibility, $\overleftrightarrow{\chi}_\parallel^{3D} d = \overleftrightarrow{\chi}_\parallel^{2D} h$, according to a parallel capacitor model[29-31]. Here, superscripts "3D" and "2D" refer to susceptibility in the supercell and in the 2D form, $\parallel$ indicates that only *xy*-plane component can be scaled, and $d$ and $h$ are simulation supercell and 2D material thickness, respectively. We use the separation distance in the bulk structure to estimate $h$, which gives 5.5 Å. The calculated $\overleftrightarrow{\chi}_\parallel^{2D}$ are shown in Fig. 2b. Detailed electronic and phononic properties can be seen in Supplementary Note 1. We find that above the phonon dispersion region (~8 THz, corresponds to 0.033 eV) and below the direct bandgap (1.6 eV), the $\chi_{11}^{el} = 25.2 > \chi_{22}^{el} = 7.8$. This can be understood from anisotropic electronic transition strength, reflected by the imaginary part of susceptibility at the direct bandgap, which determines the absorbance. The absorbance is calculated by $A_{ii}(\omega) = 1 - exp(-\omega \chi_{ii}'' d/c)$, where $c$ is the speed of light and $\chi''$ is imaginary part of susceptibility. The absorbance for the *x*-polarized and *y*-polarized light at 1.6 eV are 0.63% and 6.7%, respectively, owing to the saddle-like exciton feature and large anisotropic joint density of states. This is consistent with previous works[26], and quantitative differences come from different formulae.

In the phonon frequency region, the TO phonon modes interact with LPTL strongly. The optical branches of vibrational modes at the Γ-point can be decomposed according to the irreducible representations as

$$\Gamma_{op.} = 3A_1 \oplus 2A_2 \oplus 3B_1 \oplus 1B_2. \tag{6}$$

We find one absorbance peak $[A_{11}(\omega = 5.38\ \text{THz}) = 0.07\%]$ for the *x*-LPTL, and two absorbance peaks $[A_{22}(\omega = 1.2\ \text{THz}) = 0.26\%$ and $A_{22}(\omega = 1.87\ \text{THz}) = 0.11\%]$ for the *y*-LPTL. According to Kramers-Kronig relation, these modes could produce a macroscopic polarization and contribute to nonzero (real part) susceptibility components. Since the *y*-absorbance peaks are stronger and lower in frequency than that of the *x*-absorbance peak, the corresponding real part of $\chi_{22}^{ph}$ is also much larger than $\chi_{11}^{ph}$ at the low frequency region. For example, at the frequency of 1 THz, the real part of $\chi_{22}^{ph} = 713.4$ and $\chi_{11}^{ph} = 3.3$.

According to Eqs. (1) and (2), when an *x*-LPTL (*y*-LPTL) is applied, the FE0 phase would have largest (lowest) GFE density among the three symmetry equivalent phases. For example, on a FE0 phase one shines a LPTL ($\omega$ = 1 THz) along 60° (or 120°) with respect to the *x*-axis, then the FE0 could transit to FE1 (or FE2) phase, aligning the armchair direction parallel to the light polarization. We plot energy difference versus polarization angle in Supplementary Fig. 5. For example, if the alternative electric field strength $E_{\max}$ is chosen to be 0.2 V/nm (corresponding to an intermediate LPTL intensity of 5×10$^9$ W/cm$^2$, which is easily accessible experimentally) and frequency is 1 THz, according to Eqs. (1) and (2), the Gibbs free energy difference between the FE1/FE2 and FE0 under *x*- or *y*-LPTL illumination will be 10.66 meV/f.u. (= 1.6 μJ/cm$^2$) (Supplementary Note 5). Thus, using LPTL, one could drive a ferroic order transition schematically plotted in Fig. 1. This energy difference is proportional to the $E^2$ (or the intensity of LPTL). Increasing light intensity could boost transition kinetics.

**Anisotropic electron-energy-loss spectroscopy.** The different ferroic order geometries can be detected via ultrahigh resolution electron-energy-loss spectroscopy (EELS) in the (scanning) transmission electron microscope, especially for the vibrational signatures[32-34]. Here, we will show that this approach can be employed to distinguish different phases, owing to their anisotropic optical response. A simple approximation to describe the experimental setup is sketched inset of Fig. 3: when an electron travels along a rectilinear trajectory (velocity $v$) parallel to the material surface (with a finite distance *b*). The local dielectric tensor ($\varepsilon = 1 + \chi$) is the key component entering the classical description of electron scattering by a surface. According to a classical and quasistatic approach (Ref. 35-38 and Supplementary Note 6), energy loss probability per unit angular frequency and per unit electron path length is

$$\frac{d^2 P(\omega,b)}{d\omega dr_\parallel} = \frac{e^2}{(2\pi)^2 v^2 \varepsilon_0 \hbar} \times \int_{-\infty}^{+\infty} \text{Im}\left[\frac{\alpha(\omega,q_\parallel,q_\perp)(e^{2Qh}-1)}{e^{2Qh}-\alpha^2(\omega,q_\parallel,q_\perp)}\right] \frac{e^{-2Qb}}{Q} dq_\perp. \quad (7)$$

Here, $q_\parallel$ and $q_\perp$ are wavevectors parallel and perpendicular to the electron movement direction ($q_\parallel = \omega/v$), respectively, $Q^2 = q_\parallel^2 + q_\perp^2$, and

$$\alpha(\omega, q_\parallel, q_\perp) = \frac{\sqrt{\varepsilon_{33}(\varepsilon_\parallel q_\parallel^2 + \varepsilon_\perp q_\perp^2)/Q} - 1}{\sqrt{\varepsilon_{33}(\varepsilon_\parallel q_\parallel^2 + \varepsilon_\perp q_\perp^2)/Q} + 1} \tag{8}$$

is angle-dependent polarizability. When the system is isotropic, the polarizability reduces to its well-known form $\alpha = \frac{\varepsilon-1}{\varepsilon+1}$. We plot the EELS spectra when the electron is moving along the *x*- and *y*-direction (Fig. 3). One could clearly observe large direction dependent EELS feature, especially in the phonon region. This anisotropic vibrational EELS originates from different infrared-active phonon characters of the *x*- and *y*-LPTL in the *β*-GeSe monolayer. This result provides a high resolution damage-free approach to distinguish the geometric structure and anisotropy when the ferroic order switches.

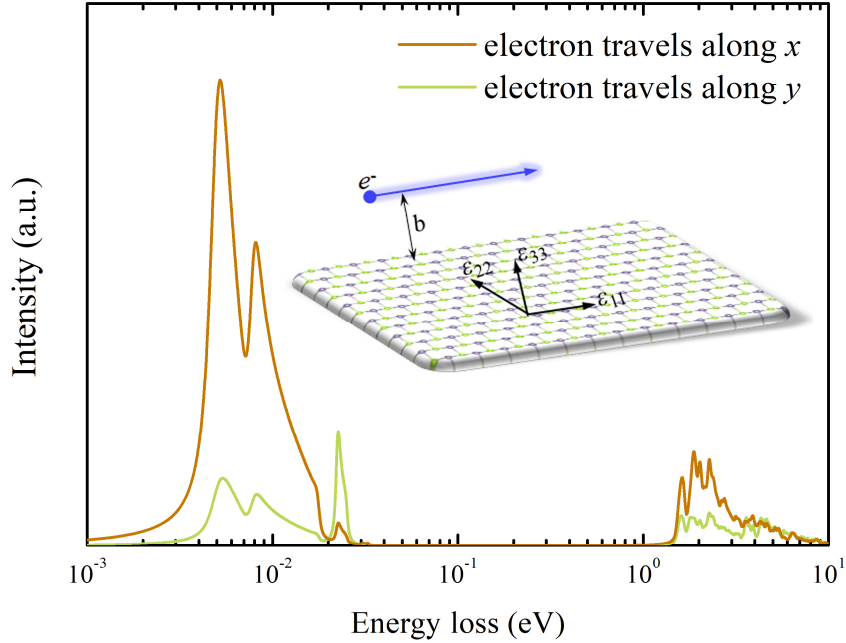

**Fig. 3 Direction dependent EELS spectra of FE0-*β*-GeSe.** The 65 keV electron beam is travelling at a distance of $b = 25$ nm parallel to the *β*-GeSe plane, along *x*- and *y*-directions (inset).

**Monolayer *α*-SnTe.** The monolayer *β*-GeSe has three orientation variants with 120° rotation to each other, owing to the $\hat{C}_3$ character of the parental geometry. Thus, even though the LPTL response is largely anisotropic (~66 times difference in the *x*- and *y*-directions at incident frequency of 1 THz), the energies separating different orientation variants are on the order of 1 μJ/cm². Now we consider another 2D time-reversal invariant multiferroic material, monolayer *α*-SnTe, whose parental geometry is $C_4$ symmetric[39,40]. As shown in Fig. 4(a), the *α*-SnTe also shows a puckered structure, with slight expansion (compression) along the *y*- (*x*-)direction. Note that even though bulk and multilayered *α*-phase other group IV-VI compounds (such as *α*-GeS, *α*-GeSe, *α*-SnS, and *α*-SnSe) have been experimentally seen, their monolayer remains to be fabricated. Hence, we only focus on the *α*-SnTe monolayer, and similar results for analogues can be obtained. Our relaxation reveals that the structure also belongs to be the *Pmn*2$_1$ space group, and the deformation strain tensor of this ferroelastic structure (FE1) is $\eta_1 = \begin{pmatrix} -0.011 & 0 \\ 0 & 0.011 \end{pmatrix}$, and the other ferroelastic structure (FE2) is subject to 90°-rotation with $\eta_2 = \begin{pmatrix} 0.011 & 0 \\ 0 & -0.011 \end{pmatrix}$. According to the modern theory of polarization, we find that its spontaneous polarization is 0.13 nC/m, along the puckered direction. These results agree well with previous works[39,40].

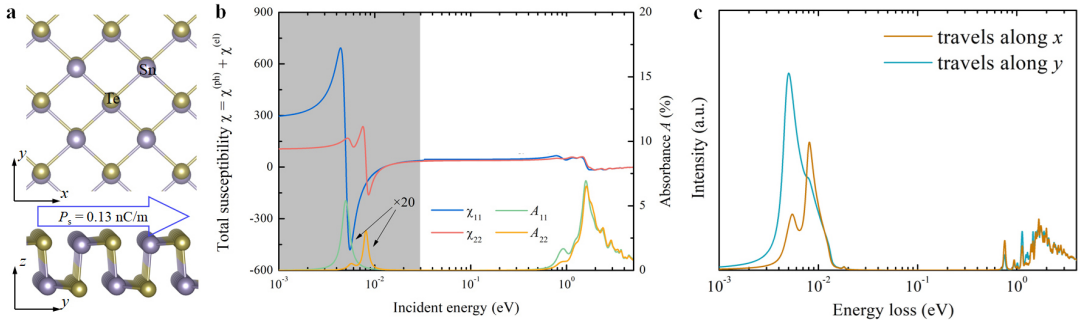

**Fig. 4 Atomic and THz optical responses of monolayer *α*-SnTe.** (a) Geometric structure of FE1-*α*-SnTe monolayer. FE2 is 90°-rotated. (b) Calculated optical susceptibility and absorbance function with respect to incident energy. The gray shaded region indicates phononic range. The phononic absorbance is enlarged for clarity reasons. (c) EELS spectra when an electron beam with kinetic energy of 65 keV travels along the *x*- and *y*-direction, at a distance of 25 nm from the α-SnTe plane.

We employ DFT and DFPT methods to calculate the electron and phonon dispersions (Supplementary Note 1), and compute the optical susceptibility (Fig. 4b). We find the electronic contributed optical response anisotropy in *α*-SnTe monolayer is not as large as that in the *β*-GeSe. At the direct optical bandgap (0.9 eV), the absorbances of the *x*-polarized and *y*-polarized light are 0.9% and 1.6%, respectively. The electronic contributed real part of susceptibility is $\chi^{el}_{11} = 44.7 > \chi^{el}_{22} = 38.3$ below the direct optical bandgap. The anisotropic excitonic absorption at different wavelength have been reported for other multiferroic 2D group-IV monochalcogenide monolayers[41], which should yield polarization direction dependent imaginary part of optical susceptibility. Then the anisotropic electron contributed real part of susceptibility at low frequency can be obtained, according to Kramers-Kronig relation. As for the phononic contribution, there is one clear absorbance peak for the *x*-LPTL (at 1.2 THz, $A_{11}$ = 0.3%), and two peaks for the *y*-LPTL (at 1.4 and 1.9 THz, with $A_{22}$ = 0.03% and 0.2%, respectively). Hence, the phononic contributed real part of susceptibility also possesses a large anisotropy. In total, when the incident energy is 0.004 eV (1.1 THz), the $\chi_{11}$ = 731.5 > $\chi_{22}$ = 137.4. For example, if a *y*-LPTL (with 1.1 THz frequency and 5×10$^9$ W/cm$^2$ intensity), the FE1 orientation variant (as shown in Fig. 4a) would have higher GFE density than the FE2 orientation variant by 12.3 meV/f.u. (3.7 μJ/cm$^2$). When one increases the LPTL intensity to 2.2×10$^{10}$ W/cm$^2$ (0.42 V/nm), this ferroic order switch (FE1 to FE2) can be *barrier-free* (Supplementary Note 7). A similar *x*-LPTL can drive the FE2 to FE1 switch, suggesting its good reversibility. Such barrierless phase transition does not require latent heat and can occur anywhere LPTL is shined. This indicates a spinodal-decomposition in the reaction coordinate space, avoiding the conventional nucleation-and-growth kinetics. Such process only requires one or a few vibrational oscillatory process, which is ultrafast and could occur on the order of several picoseconds[42].

Note that this system is interesting as its spontaneous polarization favors to align perpendicular to the optical polarization direction. This is even correct when the LPTL frequency reduces to zero – a static electric field. Since $\chi_{11}(\omega = 0) > \chi_{22}(\omega = 0)$, under *x*-directional electric field (magnitude larger than 0.3 V/nm), the spontaneous polarization prefers to align along *y*, counterintuitive with the conventional **E**//**P** picture.

The direction dependent EELS spectra is also evaluated (Fig. 4c). We again observe that the vibrational EELS shows a large anisotropy. The main EELS spectra peak positions differ by about 1 THz when the electron is moving along the *x*- (perpendicular to spontaneous polarization $\vec{P}_s$) and *y*- (parallel to $\vec{P}_s$) directions. This signal shows larger directional contrast than that in the *β*-GeSe monolayer, boosting an ultrahigh resolution characterization of ferroic orders of α-SnTe monolayer.

**Direction dependent second harmonic generation response.** We calculate the electronic contribution of SHG susceptibility $\chi^{(2)}_{abc}(-2\omega;\omega,\omega)$ of FE0-β-GeSe and FE1-α-SnTe, where $a, b, c$ are Cartesian coordinates, to measure the ferroicity via nonlinear optical response. Under the electric field with angular frequency of $\omega$, the second order nonlinear polarization takes the form $P_a(2\omega) = \chi_{abc}(-2\omega;\omega,\omega)\mathcal{E}_b(\omega)\mathcal{E}_c(\omega)$, which is correlated with SHG emitted field. The SHG susceptibility can be expressed as[43-45]

$$\chi^{(2)}_{abc}(-2\omega;\omega,\omega) = \chi^{II}_{abc}(-2\omega;\omega,\omega) + \eta^{II}_{abc}(-2\omega;\omega,\omega) + \sigma^{II}_{abc}(-2\omega;\omega,\omega), \quad (9)$$

where $\chi^{II}_{abc}(-2\omega;\omega,\omega)$, $\eta^{II}_{abc}(-2\omega;\omega,\omega)$, and $\sigma^{II}_{abc}(-2\omega;\omega,\omega)$ are interband transition, modulation of the linear susceptibility due to intraband motions of electrons, and modification by the polarization energy associated with interband motions, respectively. Specially, they can be evaluated by

$$\chi^{II}_{abc}(-2\omega;\omega,\omega) = \frac{e^3}{\hbar^2}\sum_{nml}\int\frac{d^3k}{(2\pi)^3}\frac{r^a_{nm}r^b_{ml}r^c_{ln}}{\omega_{ln}-\omega_{ml}}\times\left(\frac{f_{ml}}{\omega_{ml}-\omega}+\frac{f_{ln}}{\omega_{ln}-\omega}+\frac{2f_{nm}}{\omega_{mn}-2\omega}\right), \quad (10)$$

$$\eta^{II}_{abc}(-2\omega;\omega,\omega) = \frac{e^3}{\hbar^2}\int\frac{d^3k}{(2\pi)^3}\left\{\sum_{nml}\omega_{mn}r^a_{nm}\{r^b_{ml}r^c_{ln}\}\left[\frac{f_{nl}}{\omega^2_{ln}(\omega^2_{ln}-\omega)}+\frac{f_{lm}}{\omega^2_{ml}(\omega^2_{ml}-\omega)}\right]-\right.$$

$$\left.8i\sum_{nm}\frac{f_{nm}r^a_{nm}}{\omega^2_{mn}(\omega_{mn}-2\omega)}\{\Delta^b_{mn}r^c_{mn}\}-2\frac{\sum_{nml}f_{nm}r^a_{nm}\{r^b_{ml}r^c_{ln}\}(\omega_{ln}-\omega_{ml})}{\omega^2_{mn}(\omega_{mn}-2\omega)}\right\}, \quad (11)$$

$$\sigma^{II}_{abc}(-2\omega;\omega,\omega) = \frac{e^3}{2\hbar^2}\int\frac{d^3k}{(2\pi)^3}\left\{\sum_{nml}\frac{f_{nm}}{\omega^2_{mn}(\omega_{mn}-\omega)}[\omega_{nl}r^a_{lm}\{r^b_{mn}r^c_{nl}\}-\right.$$

$$\left.\omega_{lm}r^a_{nl}\{r^b_{lm}r^c_{mn}\}]+i\sum_{nm}\frac{f_{nm}r^a_{nm}\{\Delta^b_{mn}r^c_{mn}\}}{\omega^2_{mn}(\omega_{mn}-\omega)}\right\}, \quad (12)$$

where the position matrix element is defined as $r^a_{nm}(k) = \frac{\langle nk|\hat{v}^a|mk\rangle}{i\omega_{nm}}$ $(n\neq m)$, the energy and Fermi-Dirac distribution difference between state-$n$ and state-$m$ at $\boldsymbol{k}$ being $\omega_{nm}(k) = \omega_n(k) - \omega_m(k)$, $f_{nm}(k) = f_n(k) - f_m(k)$. In addition, $\{r^b_{ml}r^c_{ln}\} \doteq \frac{1}{2}(r^b_{nl}r^a_{ln} + r^c_{ml}r^b_{ln})$ and $\Delta^b_{mn} = v^b_{mm} - v^b_{nn}$ is velocity difference. Here the clear dependence on wavevector $\boldsymbol{k}$ is omitted. We fit the DFT calculated electronic structure with maximally localized Wannier functions[46] and use a dense $\boldsymbol{k}$-mesh sampling of (81×81×1) to perform the integration. Once the bands around the Fermi levels are sufficiently fitted, this localized basis set should yield very similar results as directly using DFT wavefunctions[47]. Note that when more bands are included, the better converged results could be obtained. In our calculations, we fit 32 bands by including all s and p orbitals of Ge, Se, Sn, and Te atoms, which could reproduce the bands between –5 and 5 eV around the Fermi level. We then assume that those band set can serve as a good approximate to a full complete wavefunction basis set and perform calculations. We calculate the $\chi_{yxx}(-2\omega;\omega,\omega)$ which reflects the *y*-polarized response under *x*-polarized light, as shown in Fig. 5. The previously mentioned re-scaling scheme is also applied. The real and imaginary parts of $\chi_{yxx}(-2\omega;\omega,\omega)$ are shown as red and green curves (Figs. 5a and 5d), which satisfies Kramers-Kronig relationship. The first peak of GeSe lies at $\hbar\omega$ = 0.82 eV, with $|\chi_{yxx}(-2\omega;\omega,\omega)|$ is 0.21 (= 0.11 − 0.17×*i*) nm/V (Fig. 5b). The momentum distribution of SHG is shown in the inset. One clearly observes that it is mainly contributed around the R (=1/2, 1/2, 0) point, consistent with its direct electronic bandgap at R (Supplementary Note 1). As for the SnTe, its smaller bandgap yields larger SHG susceptibility. At $\hbar\omega$ = 0.36 eV, $|\chi_{yxx}(-2\omega;\omega,\omega)|$ becomes as large as 24.4 (= 14.54 + 19.53×*i*) nm/V (Fig. 5e). From its momentum distribution, the dominate contribution comes from ±Λ [= (0, ±0.55, 0) Å$^{-1}$] point, which are the direct bandgap position of its electronic band structure (Supplementary Note 1). One also notes that $\chi_{xxx}(-2\omega;\omega,\omega) = \chi_{yxy}(-2\omega;\omega,\omega) = \chi_{yyx}(-2\omega;\omega,\omega) = 0$ due to mirror symmetry $\mathcal{M}_x$. We calculate other components (Supplementary Note 8) and plot polar distribution of anisotropic SHG susceptibility $|\chi_\parallel(\theta)|$ (red curve) and $|\chi_\perp(\theta)|$ (green curve) in Figs. 5c and 5f, where $\theta$ is angle that between the light polarization and *x*-axis. This also corresponds to ferroicity dependent SHG susceptibility. For example, for 90°-rotated FE2-*α*-SnTe, the $\chi_{xxx}(-2\omega;\omega,\omega) \neq 0$ and $\chi_{yxx}(-2\omega;\omega,\omega) = 0$.

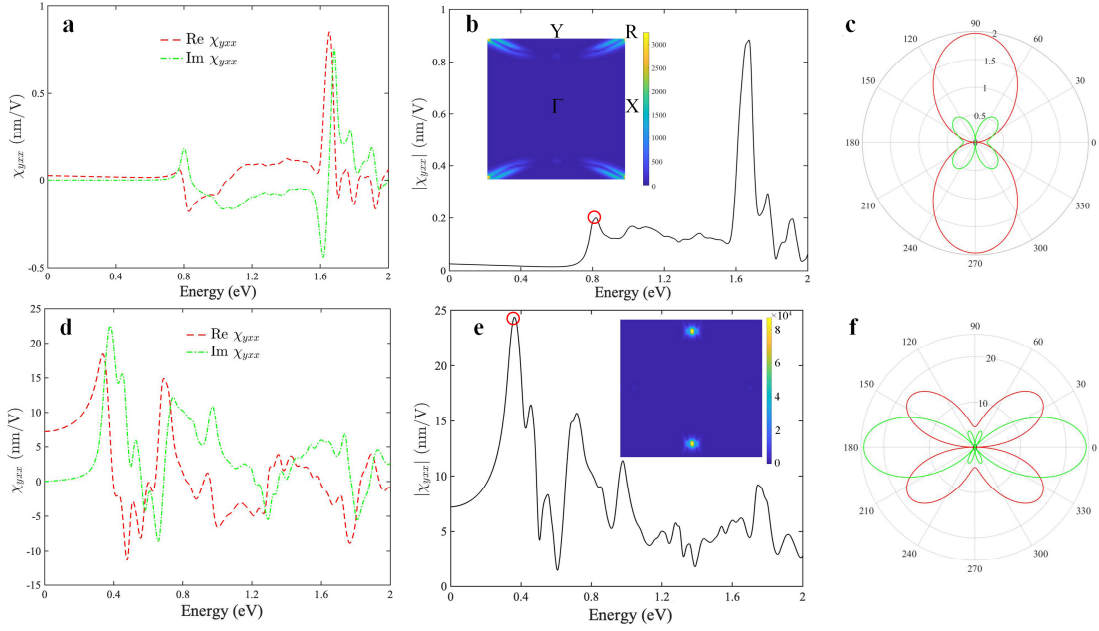

**Fig. 5 DFT calculated second harmonic generation results.** Real part, imaginary part, and total frequency dependent in-plane electronic contributed SHG spectra of monolayers (a,b) FE0-$\beta$-GeSe and (d,e) FE1-$\alpha$-SnTe. Horizontal axis indicates incident energy $\hbar\omega$. Inset shows ***k***-resolved total SHG susceptibility distribution at $\hbar\omega$ = 0.82 eV (GeSe) and 0.36 eV (SnTe), indicated by the red circles. Polarization anisotropy of absolute SHG susceptibilities of (c) GeSe and (f) SnTe, where red and green curves indicate SHG susceptibility that parallel and perpendicular to the polarization direction, respectively.

We briefly compare our SHG calculations with previous works on other centrosymmetry broken 2D time-reversal multiferroic materials. The point group of both $\beta$-GeSe and $\alpha$-SnTe are both $C_{2v}$, which suggests that their SHG anisotropy is the same as those of other $\alpha$-phase group IV-VI monolayers, such as monolayers GeS and SnSe. It is well-known that the standard DFT calculations underestimates the electronic bandgap, hence in principles one has to adopt more accurate approaches, such as hybrid functional or many body calculations[48]. Unfortunately, this is extremely computational challenging because of nonlocal interactions included and very dense ***k***-mesh needed. In the previous calculations[47], Wang and Qian adopted a scissor operator scheme to shift the DFT calculated bandgap to be consistent with the optical bandgap calculated through many body theory with exciton interaction correction. In our current study, we only focus on the relative strength of SHG responses and its anisotropy, and the scissor correction is not applied here. Therefore, in experiments the peaks of the SHG responses would shift to higher energy compared with our calculation results, usually by ~ 0.5 – 1 eV. According to Eqs. (10)–(12), the SHG peak magnitude roughly scales as $1/E_g$ ($E_g$ is the electronic bandgap). Hence, the experimentally measured SHG peak values would also be smaller than our calculation results. Nevertheless, the overall SHG shape and anisotropy trend should be similar with our calculations.

**Discussion.** In our current LPTL driven phase transition mechanism, we only elaborate the real part of optical susceptibility and assumes its imaginary part to be small. Actually when the optical frequency is chosen around finite Im($\chi_{ii}$), direct optical absorption occurs. This corresponds to a conversion from photons to infra-red phonon or electron-hole pairs (electron interband excitation). Actually this could also trigger mechanical strains[49] to the systems, and phase transition may occur[50,51]. However, they would generate unwanted heat into the system via phonon-phonon interaction or non-radiative electron-hole recombination so that the device reversibility will reduce. In our mechanism, only electric field work done is applied to the systems. Even though all of the work done converts to be internal energy, the temperature rise is still very small[30]. Hence, in the current discussion, we only focus on the real part of optical susceptibility contribution and avoid such direct photon absorption process.

In addition, we note that in the current setup only ferroelastic order degeneracy can be broken under LPTL irradiation. One could not break the degeneracy between the ferroelectric phases with opposite polarization states ($\vec{P}_s$ and $-\vec{P}_s$). Thus, one needs to apply additional fields (such as zero frequency static electric field or introducing surface/interface effects) to break such degeneracy. Actually, optical control of the polarization direction (reversal by 180°) has been observed experimentally in BaTiO$_3$ thin films deposited on transition metal dichalcogenide monolayers[52], where surface effects can be observed. This suggests that when additional spatial broken interactions exist, one could further break the polarization degeneracy and control/manipulate the ferroelectric order.

In the current model, we use a single unit cell to perform calculations. This indicates a coherent phase transition, with all unit cells rotate their ferroic orders simultaneously. In reality, the materials could contain domain walls, which spatially separate different ferroic orientation variants. The phase transition usually occurs along with the domain wall movement[2]. Compared with 3D materials or thick films (such as $BaTiO_3$ perovskite) in which the domain wall is a 2D interface, in 2D materials, the domain wall is actually in quasi-1D. This would make the phase transition in 2D materials much easier than the 3D materials or thick films, with smaller residue stress during transition. Hence, the phase transition would cost low energy and can be nonvolatile. Previous works have shown that the domain wall migration energy is around a few to a few tens meV/Å[28,41], much smaller than the formation energy of 1D defects in 3D materials, indicating a fast domain wall assisted order switching. Therefore, the existence of domain wall may facilitate the phase transition. In addition, light usually has a large spot size, on the order of μm or larger. Once the light polarization, intensity, and frequency are carefully selected, it can trigger ultrafast and barrier-free phase transition wherever it irradiates. The conventional nucleation and growth kinetics can be avoided, and a coherent phase transition can be expected.

However, if the material contains a large number of point, line or area defects, they may strongly affect the optical response functions, by siginificantly reducing the carrier lifetime ($\tau$) and introducing doping levels into the phonon and electron structures. For example, defects usually brings both shallow and deep energy levels into semiconductor electronic bands[53]. Then the low frequency light may be absorbed and the imaginary part of optical susceptibility needs to be considered. When the defect or impurity concentration is not high enough, their effects can be phenomenologically incorporated by the finite lifetime in the response formulae, in Eqs. (4) and (5). In order to avoid such uncertainty, we choose the light frequency away from resonant absorption frequencies, then the exact value of lifetime does not affect the estimate too much. However, an exact and complete evaluation is very complicated, and is out of the scope of our current study.

The phase transition related strain in the current study is within 8%, which is usually sustainable for 2D materials. However, one may notice that if the sample size is a few to a few tens of nanometers, then such strain would cause a big lattice mismatch in the system, and may even affect the chemical bonds at the boundary between the transformed and untransformed domains. A direct numerical simulation of such process in a large area is very computationally challenging and memory demanding. Actually one may allow a freestanding 2D materials to be slightly slack in the *z*-direction, or carefully select a surface to support them with weak interactions (such as van der Waals)[54]. This allows the atoms to move in the *z*-direction with small energy cost, which could effectively release the in-plane strains during martensitic phase transitions. This is different from 3D bulk materials or thick films, where such strain induced damage can be significantly large and is fatal to their reversible usage.

**Conclusion.** In conclusion, we implement a theoretically and computationally combined approach to study the electronic, phononic, and mechanical responses of 2D time-reversal invariant multiferroic monolayers under LPTL illumination. Taking two experimentally fabricated multiferroic *β*-GeSe and *α*-SnTe monolayers as examples, we find that they both show a large anisotropic optical response, especially at the terahertz region, owing to their selective vibrational infrared-active vibrational modes. According to the thermodynamic theory, we predict that LPTL can efficiently drive phase transition with an ultrafast kinetics (or even a barrierless GFE profile). This noncontacting optomechanical approach to switch the ferroelastic order of 2D materials can be easily controlled experimentally. In order to detect different orders, we propose to measure vibrational EELS spectra, which is direction dependent and has an ultrahigh resolution experimentally. Anisotropic SHG response calculations are also provided. Different from the parameterized phonon-phonon coupling models in the optically induced phase transition, we provide a first-principles quantitative estimation on the terahertz optics effects. Such mechanism can also apply to other frequency range, as long as the direct optical absorption is eliminated. Owing to the rapid developments of using terahertz laser triggering (topological or structural) phase transition, our theory provides a route to explain these experiments and predict unexplored phenomena from a precise first-principles approach.

**Methods.** Our first-principles calculations are based on density functional theory[14] and performed in the Vienna *Ab initio* Simulation Package (VASP)[15] with generalized gradient approximation (GGA) treatment of exchange and correlation functional in the solid state PBE form (PBEsol)[16]. In order to simulate two-dimensional (2D) materials, a vacuum distance of 15 Å in the out-of-plane $z$-direction is applied to eliminate layer interactions. The projector augmented wave method[17] and planewave basis set are applied to treat the core and valence electrons, respectively. The kinetic cutoff energy of planewave is set to be 350 eV. The reciprocal space is represented by Monkhorst-Pack $k$-mesh scheme[18]. The electron and force component convergence criteria are set to be $10^{-7}$ eV and 0.001 eV/Å, respectively. Spin-orbit coupling (SOC) interactions are included self-consistently. For the SnTe monolayer, $d$-electrons are incorporated in the Sn valence electrons, and Grimme's D3 scheme[19] is used to include van der Waals interactions, which has been demonstrated to yield good results on the energy barrier[39]. We use modern theory of polarization based on Berry phase approach[55,56] to evaluate spontaneous polarization (Supplementary Note 2). The optical dielectric function contributed from the electron subsystems are calculated according to random phase approximation (RPA). In order to evaluate dielectric function contributed from the ionic subsystem (phonons), we use density functional perturbation theory (DFPT)[20] to calculate vibrational modes and Born effective charge of each ion, with the help of the Phonopy code[21].

**Acknowledgement.** This work is supported under the National Key Research and Development Program (Grant No. 2019YFA0210600), the National Natural Science Foundation of China (NSFC) under Grant Nos. 21903063, 11974270, and 11904350, and the Young Talent Startup Program of Xi'an Jiaotong University. S. Z. also acknowledges the support of Anhui Provincial Natural Science Foundation (Grant. No. 2008085QA30).

**Corresponding authors**: *J.Z.: jianzhou@xjtu.edu.cn; †S.Z.: szhang2@ustc.edu.cn

**Competing Interests.** The authors declare no competing interests.

**Author Contribution.** J.Z. conceived the concept. J.Z. and S.Z. performed calculations. J.Z. analyzed data. J.Z. and S.Z. wrote the manuscript.

**Data Availability.** All data generated or analyzed during this study are included in this published article (and its Supplementary Information files), and are available from the authors upon reasonable request.

**Code Availability.** The related codes are available from the authors upon reasonable request.

# Supplementary Information for "Terahertz Optics Driven Phase Transition in Two-Dimensional Multiferroics"

Jian Zhou[1,*], Shunhong Zhang[2,†]

[1] *Center for Advancing Materials Performance from the Nanoscale, State Key Laboratory for Mechanical Behavior of Materials, Xi'an Jiaotong University, Xi'an, 710049, China*

[2] *International Center for Quantum Design of Functional Materials (ICQD), Hefei National Laboratory for Physical Sciences at Microscale, and CAS Center For Excellence in Quantum Information and Quantum Physics, University of Science and Technology of China, Hefei, Anhui 230026, China*

Emails: [*] jianzhou@xjtu.edu.cn; [†] szhang2@ustc.edu.cn

# I. Supplementary Note 1: Electronic and phonon dispersions and contributed dielectric functions.

We plot the electronic band structure and phonon dispersion along the high symmetric *k*-path, of both FE0-*β*-GeSe and FE1-*α*-SnTe monolayer (Supplementary Fig. 1). One can see that the *β*-GeSe monolayer is an indirect bandgap semiconductor, with its indirect and direct bandgap values of 1.5 and 1.6 eV, respectively. The *α*-SnTe monolayer is also an indirect bandgap semiconductor, with its indirect and direct bandgaps are 0.7 and 0.8 eV, respectively, similar as *α*-SnSe monolayer[1]. The irreducible representation analyses of low energy band states reveal the anisotropic electronic absorption (along *x*- and *y*-directions) of them. Hence, the electronic absorbance spectra is direction-dependent. This demonstrates the anisotropy of electronic contributed susceptibility below the bandgap, according to Kramers-Kronig relationship.

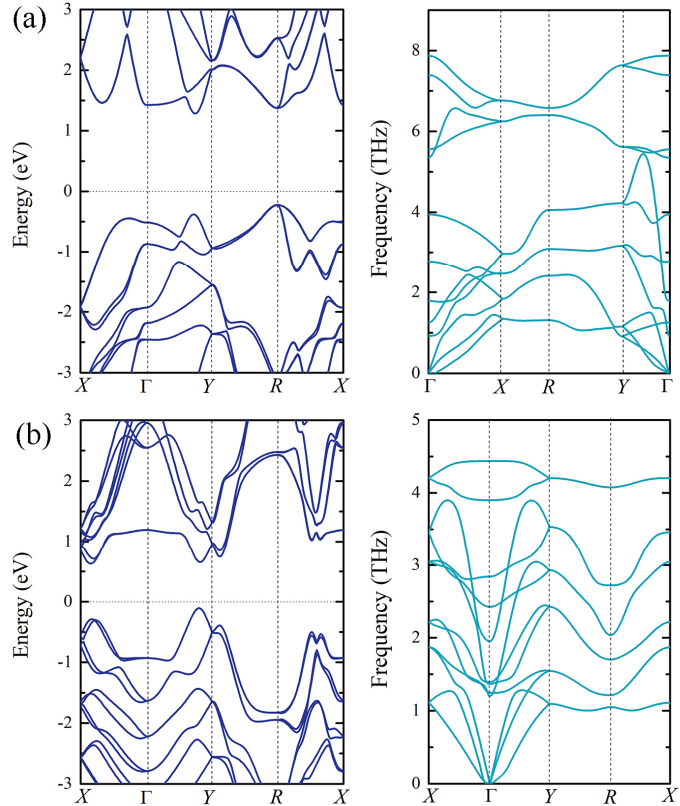

**Supplementary Figure 1.** Electronic and phonon dispersions of (a) FE0-*β*-GeSe and (b) FE1-*α*-SnTe. SOC is included.

From phonon dispersions, we observe no imaginary frequencies over the whole Brillouin zone (BZ). This suggests their dynamic stability, and confirms that they can exist in free-standing monolayer forms. In addition, we plot the mass averaged vibrational modes at the Γ-point that contributed to the phononic susceptibility (Supplementary Figs. 2 and 3). One sees the infra-red active feature of these modes.

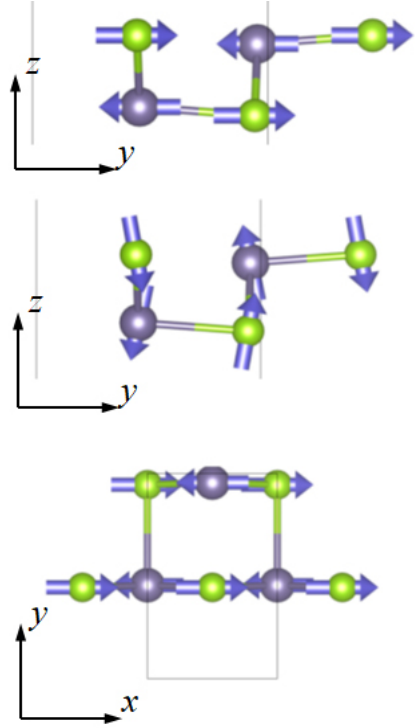

**Supplementary Figure 2.** Three main vibrational modes that contribute to $\chi_{yy}$ (up and middle panels, at 1.2 THz and 1.87 THz) and $\chi_{xx}$ (bottom panel, at 5.38 THz) of GeSe.

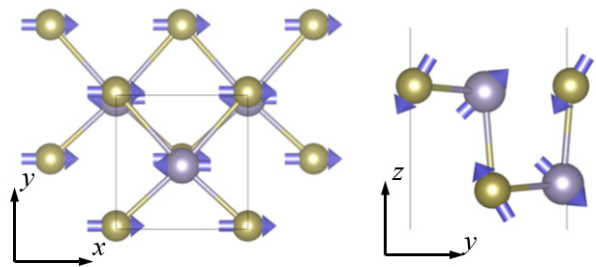

**Supplementary Figure 3.** Two main vibrational modes that contribute to $\chi_{xx}$ (left panel) and $\chi_{yy}$ (right panel) of SnTe.

**II. Supplementary Note 2: Calculation details of spontaneous polarization**

According to the modern theory of polarization, the exact electric polarization value is not well-defined, and one could only evaluate the polarization change according to the Berry phase approach, which measures the spontaneous polarization from a centrosymmetric structure

$$\mathbf{P}_s = \mathbf{P}^{\text{f}} - \mathbf{P}^{\text{i}} = \frac{1}{S}\sum_j \left(q_j^{\text{f}} \mathbf{r}_j^{\text{f}} - q_j^{\text{i}} \mathbf{r}_j^{\text{i}}\right)$$

$$- \frac{i|e|}{(2\pi)^2} \sum_{n \in \text{occ}} \int_{\text{BZ}} d^2\mathbf{k} \times \left[\langle u_{n\mathbf{k}}^{\text{f}}|\nabla_\mathbf{k}|u_{n\mathbf{k}}^{\text{f}}\rangle - \langle u_{n\mathbf{k}}^{\text{i}}|\nabla_\mathbf{k}|u_{n\mathbf{k}}^{\text{i}}\rangle\right] \quad (1)$$

Here we use the 2D Brillouin zone, and the scripts 'i' and 'f' refers to the initial and finial structures, respectively. The first term incorporates ionic contributions to the polarization, where $S$ is the area of a simulation unit cell and $j$ runs for all ions. The second term is the difference of Berry phase and the $|u_{n\mathbf{k}}\rangle$ is the cell periodic part of Bloch wave function of band-$n$ at $\mathbf{k}$. In this expression, the polarization could shift by a quanta $\Delta \mathbf{P} = e\mathbf{R}/S$, where $\mathbf{R}$ is the lattice vector. In practice, we choose the conventional strategy to keep the centrosymmetric system to be non-polarized ($\mathbf{P} = 0$), and displace the ions to the energetic ground state. During displacement, we allow no ions to move across simulation unit cell, of which the origin is chosen to be the reference point to measure the distance $\mathbf{r}$. In this case, the polarization direction would point from the anions to cations in one unit cell, which agrees with the conventional concept of dipole moment for a finite system.

## III. Supplementary Note 3: Spontaneous polarization flip contributing to the Gibbs free energy

Under THz irradiation, when the frequency of light is lower than the Debye vibration frequency of polarization reversal and the electric field is strong enough, the polarization can flip back and forth ($P_s$ to $-P_s$) under the electric field. Hence, the spontaneous polarization is not time independent and could contribute to the Gibbs free energy, different from the previous works where higher frequency light is used[1]. For simplicity, we estimate the ideal coercive field by $E_c = U/(S_{\text{f.u.}} P_s)$, where $U$ indicates the energy barrier (per f.u.) that separates the polarization reversal (between $P_s$ and $-P_s$) and $S_{\text{f.u.}}$ is the area of one formula unit. One has to note that this energy barrier can be different from the energy difference between the high symmetry structure and ferroelectric ground state, and such ideal coercive field is very roughly estimated. Actually one can also estimate the atomic position change according to a simple damped driven oscillator model. Under a sinusoidal electric field, the system oscillates according to

$$\frac{d^2 x_\mu}{dt^2} + 2\zeta \omega_m \frac{dx_\mu}{dt} + \omega_m^2 x_\mu = \frac{1}{m_\mu} E_{\max} Z_\mu^* \cos \omega t \quad (2)$$

where $x_\mu$, $m_\mu$, and $Z_\mu^*$ are average values of displacement, mass, and Born effective charge of ion-$\mu$, respectively. $\omega_m$ is phonon frequency and $\zeta$ is effective damping. The displacement amplitude then can be evaluated by

$$x_\mu^{\max} = \frac{Z_\mu^* E_{\max}}{m_\mu \omega \sqrt{(2\omega_m \zeta)^2 + \left(\frac{\omega_m^2 - \omega^2}{\omega^2}\right)^2}} \quad (3)$$

Since the phonon frequency is only 1.2 THz for GeSe and 1.4 THz for SnTe, we estimate that under electric field amplitude of $E_{\max} = 0.2$ V·nm$^{-1}$ at frequency 1 THz, the ion amplitude is on the order of 1 Å, which is sufficiently large for $P_s$ and $-P_s$ reversal. This also indicates that $E_{\max}$ is much larger than $E_c$, consistent with previous analysis.

In order to estimate the polarization reversal contributed Gibbs free energy (GFE), we plot a simple polarization reversal process as in Supplementary Fig. 4. One could consider the following two steps (for simplicity, we first take 1D model):

(i) The sinusoidal electric field $\mathcal{E}(t) = E_{max} \sin \omega t$ is between $-E_c$ to $-E_{max}$ and to $E_c$ (From step ④ to ① to ②): Once an electric field $-E_c$ is applied, the system jumps to polarization $-P_s$. It contributes to the GFE to be $-S_{f.u.} P_s \mathcal{E}(t)$. Note that the linear response (which depends on susceptibility $\chi$) is not included here, which will be discussed in the main text.

(ii) The electric field between $+E_c$ to $+E_{max}$ and back to $-E_c$ (② to ③ to ④). This process is a reversal process as (i), and gives $-S_{f.u.} P_s \mathcal{E}(t)$ to GFE.

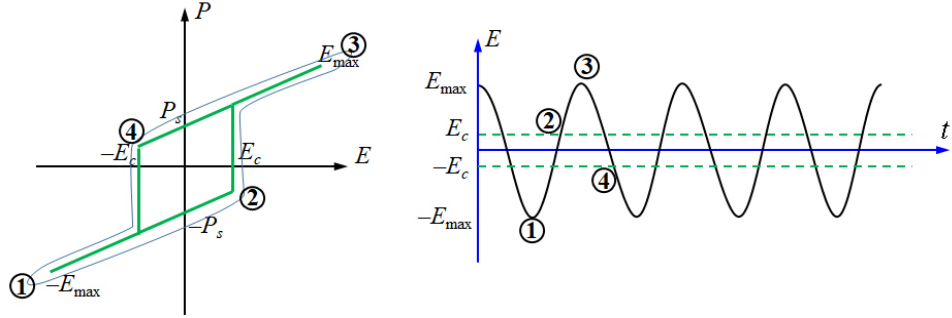

**Supplementary Figure 4.** THz optics driven polarization reversal and its contribution to Gibbs free energy. Left panel is a simplified *E-P* plot of ferroelectrics, and the sinusoidal electric field variation is plotted in the right panel. The slope of the green line in the *E-P* plot corresponds to the optical susceptibility that contributed from both electron and phonon subsystems. The two horizontal dashed lines indicates the critical electric field $\pm E_c$.

One could then integrate over time of the whole process to obtain the total GFE contribution. It is easy to see that between the two dashed horizontal lines in the right panel of Supplementary Fig. 4, the contribution to GFE cancels each other. Thus, only the $E_c \to E_{max} \to E_c$ and $-E_c \to -E_{max} \to -E_c$ processes give finite GFE, which are the same to each other. One easily obtains the results to be

$$\mathcal{g}_0 = -\frac{E_{max} P_s}{2} \sqrt{1 - \frac{E_c^2}{E_{max}^2}} \tag{4}$$

In our current case, the critical electric field (on the order of 0.01 V·nm$^{-1}$) is order of magnitude smaller than the optical field strength, thus the above equation can be approximated to be $\mathcal{g}_0 = -\frac{E_{max} P_s}{2}$. If the electric field is not parallel to the polarization direction, one can decompose the electric field into $\mathbf{E}_\parallel = \mathbf{E} \cos \theta$ and $\mathbf{E}_\perp = \mathbf{E} \sin \theta$, where $\theta \in [0, \frac{\pi}{2}]$ is the angle between the LPTL polarization and the polarization $\mathbf{P}_s$.

The above equation then becomes

$$\mathcal{G}_0 = -\frac{E_{max}P_s \cos\theta}{2} \tag{5}$$

This is the Eq. (1) in the main text. Note that if the light frequency is highly above the intrinsic vibration frequency, or the light electric field strength is smaller than the coercive field, such $\mathcal{G}_0$ term becomes zero, and the only optical susceptibility term contributes to the total GFE density.

## IV. Supplementary Note 4: Ferroelastic phases of *β*-GeSe monolayer.

We use DFT calculations to compute the different ferroelastic orders of *β*-GeSe monolayer (Supplementary Fig. 5). In order to evaluate the spontaneous transformation strains of each ferroelastic order, we use a $(2 \times 2\sqrt{3})$ supercell of the high symmetric parental structure, and all the three ferroelastic orientation variants can be derived by Peierls distortions from this parental structure. In Cartesian coordinates, the $(2 \times 2\sqrt{3})$ basis vectors of parental structure can be denoted as $\mathbf{h}_1 = 2a\hat{\mathbf{x}}$ and $\mathbf{h}_2 = 2\sqrt{3}a\hat{\mathbf{y}}$, where $\hat{\mathbf{x}}$ and $\hat{\mathbf{y}}$ are unit vectors along the *x*- and *y*-direction, and $a$ is lattice constant of a unit cell. Then, one can define a supercell structural matrix of it as $\mathbf{H}_p = \{\mathbf{h}_1, \mathbf{h}_2\} = \begin{pmatrix} 2a & 0 \\ 0 & 2\sqrt{3}a \end{pmatrix}$. Similarly, the distorted (fully relaxation using DFT calculations) orientation variants [insets of Supplementary Fig. 5(e)] would have their corresponding structural matrices (denoted as $\mathbf{H}_i$, $i = 0, 1, 2$). We can then define three transformation matrices, $\mathbf{J}_i = \mathbf{H}_i \mathbf{H}_p^{-1}$. This could allow us to evaluate transformation strain tensors according to the definition of Green-Lagrange strain tensor as[2]

$$\boldsymbol{\eta}_i = \tfrac{1}{2}(\mathbf{J}_i^{\mathrm{T}} \mathbf{J}_i - \mathbf{I}) = \tfrac{1}{2}\left[(\mathbf{H}_p^{-1})^{\mathrm{T}} \mathbf{H}_i^{\mathrm{T}} \mathbf{H}_i \mathbf{H}_p^{-1} - \mathbf{I}\right] = \begin{pmatrix} \varepsilon_{xx}^{(i)} & \varepsilon_{xy}^{(i)} \\ \varepsilon_{yx}^{(i)} & \varepsilon_{yy}^{(i)} \end{pmatrix}. \tag{6}$$

Here, $\varepsilon_{xx}^{(i)}$ and $\varepsilon_{yy}^{(i)}$ are the uniaxial tensile or compressive strains along the *x*- and *y*-directions, respectively, for the *i*-th orientation variant, and $\varepsilon_{xy}^{(i)} = \varepsilon_{yx}^{(i)}$ is the shear strain component. The calculated transformation strain tensors are $\boldsymbol{\eta}_0 = \begin{pmatrix} 0.042 & 0 \\ 0 & -0.040 \end{pmatrix}$, $\boldsymbol{\eta}_1 = \begin{pmatrix} -0.016 & -0.041 \\ -0.041 & 0.021 \end{pmatrix}$, and $\boldsymbol{\eta}_2 = \begin{pmatrix} -0.016 & 0.041 \\ 0.041 & 0.021 \end{pmatrix}$. The difference between these transformation strain tensors represents the switch between them. We thus obtain the FE0→FE1, FE0→FE2, and FE1→FE2 relative transformation strains $\boldsymbol{\eta}_0^1 = \begin{pmatrix} -0.053 & -0.041 \\ -0.041 & 0.064 \end{pmatrix}$, $\boldsymbol{\eta}_0^2 = \begin{pmatrix} -0.053 & 0.041 \\ 0.041 & 0.064 \end{pmatrix}$, and $\boldsymbol{\eta}_1^2 = \begin{pmatrix} 0.010 & -0.083 \\ -0.083 & 0.033 \end{pmatrix}$, respectively. In Supplementary Fig. 5(e), we plot the relative energies of these orientation variants as a function of relative transformation strains. In detail, the center of horizontal axis denotes the $\boldsymbol{\eta}_0$ state, and its right-hand-side (left-hand-side) direction indicates a linear increase of strain states along the $\boldsymbol{\eta}_0^1$ ($\boldsymbol{\eta}_0^2$). Vertical dashed lines (from left to right) denote the $\boldsymbol{\eta}_2$, $\boldsymbol{\eta}_0$, and $\boldsymbol{\eta}_1$ states. Each of these three curves quadratically increases as its intrinsic strain (not the residue strain with respect to parental structure) increases, indicating that these strain states are within their own elastic deformation region. In addition, under strain state of $\boldsymbol{\eta}_0$, the FE1/FE2 orientation variant is energetically higher than FE0 by 101 meV per f.u. This energy can be released by the internal coordinate shuffling. We also calculate the elastic coefficients of β-GeSe and estimate the elastic energy corresponds to the transformation strain $\boldsymbol{\eta}_0^1$. The 2D elastic coefficients (Voigt notation) of FE0 are found to be $C_{11}$ = 32.47 GPa·nm, $C_{12}$ = −0.40 GPa·nm, $C_{22}$ = 34.48 GPa·nm, and $C_{44}$ = 8.01 GPa·nm. Then the elastic energy is estimated to be 0.1 eV per f.u., consistent with DFT results in Supplementary Fig. 5. Note that due to rotation symmetry, the FE1↔FE2 switch also has an elastic energy of 0.1 eV per f.u. [but not along the horizontal axis in Supplementary Fig. 5(e)]. This indicates that from FE-0 phase and under the strain of $\boldsymbol{\eta}_0^1$ (and elastic energy of 0.1 eV per f.u.), the structure would prefer FE1 rather than FE0, since the latter has higher strain energy. Then phase transition would occur. Similar arguments can be applied for other orientation variants. If appropriate external strain is applied to the system, this could facilitate LPTL induced phase transition.

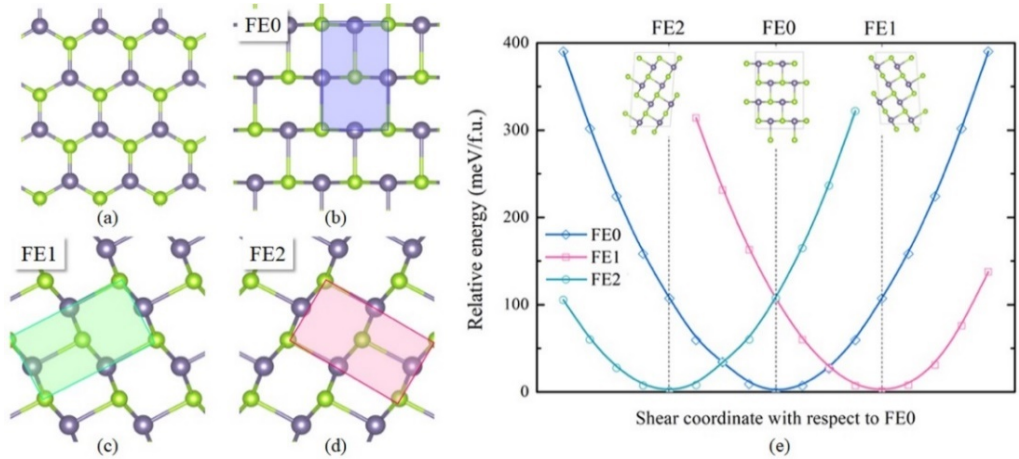

**Supplementary Figure 5.** Atomistic structure of (a) high symmetric and (b) FE0, (c) FE1, and (d) FE2 orders. Shaded areas are 120º-rotation equivalent unit cells (not simulation supercells). (e) Relative energy under shear deformation from FE0 to FE1/FE2 orders. Insets show simulation supercell cells of different phases.

## V. Supplementary Note 5: Anisotropic Gibbs free energy density of *β*-GeSe monolayer under LPTL.

We apply Eqs. (1) and (2) in the main text to compute the Gibbs free energy under LPTL illumination with different polarization angle. In Supplementary Fig. 6, we plot the polarization angle dependent Gibbs free energy change of FE0-*β*-GeSe monolayer. The angle $\theta$ is defined as between the polarization direction and the armchair direction (*y* in FE0-*β*-GeSe). One could observe that, due to small *x*-component phononic susceptibility, the Gibbs free energy reduction is higher when an *x*-polarized terahertz laser is applied ($\theta = 90°$). If the initial orientation variant is FE0, one could apply a LPTL with its polarization along $\theta$ (with $\langle\theta, \hat{\mathbf{x}}\rangle = \pm\frac{\pi}{6}$) to switch it to FE1 or FE2. Similarly, from a FE1 (FE2) orientation variant, *y*-polarized LPTL can switch it to FE0.

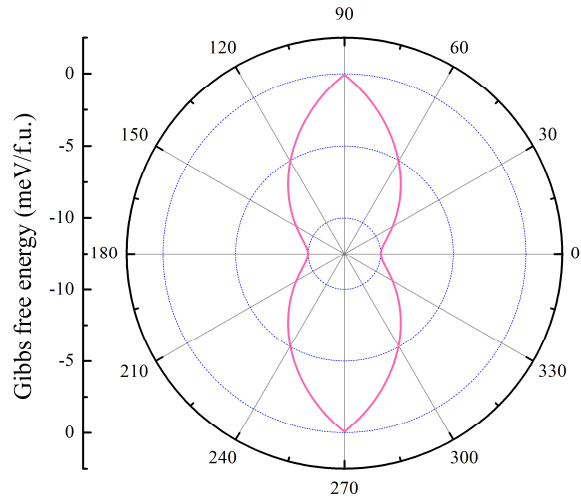

**Supplementary Figure 6.** Angle-dependent GFE change of GeSe monolayer. The laser frequency is selected to be 1 THz, and the magnitude of electric field is 0.2 V·nm$^{-1}$. The angle is defined as between the LPTL polarization direction and the GeSe armchair direction.

## VI. Supplementary Note 6: Electron energy loss spectroscopy (EELS).

EELS can provide geometry information, which uses an electron beam positioned a few tens of nanometers away from the sample ("aloof" mode). Compared with the standard transmission experiment with the same electron energy, this noninvasive and aloof mode reduces the sample damage by a factor of ~1000 (Supplementary Ref. 3). We can combine electromagnetic dynamics theory within a continuum dielectric model to obtain the EELS spectrum intensity as a function of electron energy loss for 2D materials.

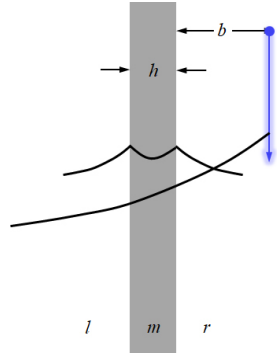

**Supplementary Figure 7.** Geometric model for fast electron travelling beside a 2D material. The normal direction to plane is denoted as $z$. Trajectory is marked by the blue arrow.

Supplementary Fig. 7 shows a simplified geometric model when a fast electron travels parallel to a 2D material surface (sandwiched by vacuum in both sides). According to non-relativistic electrodynamics, the electric potential generated by this electron can be written as (SI unit)

$$\phi_{ext} = -\frac{e}{\varepsilon_0 \varepsilon(z,q_\parallel,q_\perp)} \frac{e^{-Qb}}{Q} \delta\left(q_\parallel - \frac{\omega}{v}\right) \quad (7)$$

The $Q^2 = q_\parallel^2 + \left(\frac{\omega}{v}\right)^2$ denotes the surface excitation wavevector. From this, we can write the potential in the three regions.

$$\phi_l = -\frac{e}{Q\varepsilon_0} e^{-Q(b+h-z)} + \frac{A}{\varepsilon_0} e^{Qz} \quad (8a)$$

$$\phi_m = -\frac{e}{Q\varepsilon_0 \varepsilon_m(z,q_\parallel,q_\perp)} e^{-Q(b+h-z)} + \frac{B}{\varepsilon_0} e^{-Qz} + \frac{C}{\varepsilon_0} e^{Qz} \quad (8b)$$

$$\phi_r = -\frac{e}{Q\varepsilon_0} e^{-Q(b+h-z)} + \frac{D}{\varepsilon_0} e^{-Qz} \quad (8c)$$

Here $z < b + h$, constraining our discussion in the left side of the electron. The continuum boundary condition occurs at the boundary of these three regions, namely,

$$\phi_l(0) = \phi_m(0), \quad \phi_m(h) = \phi_r(h) \tag{9a}$$

$$\phi_l'(0) = \varepsilon_m \phi_m'(0), \quad \varepsilon_m \phi_m'(h) = \phi_r'(h) \tag{9b}$$

Using these conditions, we can determine the coefficients $A, B, C, D$. According to previous works[4,5], the anisotropic effective dielectric function satisfies $Q^2 \varepsilon_m^2 = \varepsilon_{zz}(q_\parallel^2 \varepsilon_\parallel + q_\perp^2 \varepsilon_\perp)$. This equation satisfies the Laplace's equation in the medium $\nabla(\overleftrightarrow{\varepsilon}_m \nabla \phi_m) = 0$. The energy loss probability per unit angular frequency can be defined as the work done on the electron

$$W = \int_0^{+\infty} \hbar\omega \frac{dP(\omega,b)}{d\omega} d\omega = -e \int_{-\infty}^{+\infty} \mathbf{v} \cdot \mathbf{E}(\mathbf{r}_e) \, dt \tag{10}$$

where $\mathbf{E}$ is the electric field induced by the polarizability of 2D material, and $\mathbf{r}_e$ is the electron position. From the above equation, the energy loss probability can be calculated as

$$P(\omega, b) = -\frac{e}{\hbar\omega} \int_{-\infty}^{+\infty} \frac{\partial}{\partial x_\parallel} (\phi_r - \phi_{ext}) \bigg|_{x_\parallel = vt} dx_\parallel \tag{11}$$

The EELS is measured as energy-loss probability per unit angular frequency and per unit path length

$$\frac{d^2 P(\omega,b)}{d\omega dr_\parallel} = \frac{e^2}{(2\pi)^2 v^2 \varepsilon_0 \hbar} \times \int_{-\infty}^{+\infty} \text{Im}\left[\frac{\alpha(\omega,q_\parallel,q_\perp)(e^{2Qh}-1)}{e^{2Qh} - \alpha^2(\omega,q_\parallel,q_\perp)}\right] \frac{e^{-2Qb}}{Q} dq_\perp \tag{12}$$

This is Eq. (7) in the main text. From Figs. 4 and 5 in the main text, the vibrational EELS shows clear direction-dependent feature. When the electron beam is travelling along an arbitrary direction $\theta$ (with respect to the $x$-axis), the parallel and vertical dielectric function is $\varepsilon_\parallel = \cos^2\theta \, \varepsilon_{xx} + \sin^2\theta \, \varepsilon_{yy}$ and $\varepsilon_\perp = \sin^2\theta \, \varepsilon_{xx} + \cos^2\theta \, \varepsilon_{yy}$.

**VII. Supplementary Note 7: Barrier-free ferroic order switch of *α*-SnTe monolayer.**

We use cell variable NEB to estimate the energy barrier between different orientation variants in *α*-SnTe monolayer. The energy barrier density separating the two orientation variants is 2.3 meV per f.u. Note that unlike conventional ferroic phase transition where the high symmetry non-ferroic structure serves as the transition saddle point on the energy landscape, the saddle point belongs to *Abm*$_2$. This structure possesses an in-plane polarization, pointing along the 45°-direction (with respect to the *x*-axis). We assume it still serves as transition saddle point under light and calculate its optical response functions. For the phonon contributed susceptibility, we only include the real frequency contribution, while the imaginary modes are ignored.

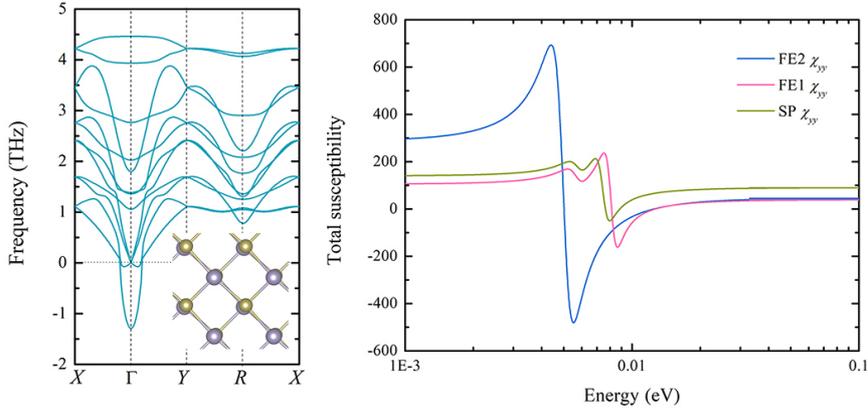

**Supplementary Figure 8.** (a) Phonon spectrum of the SP from FE1 to FE2 phase transition of *α*-SnTe. Inset shows its atomic geometry. (b) Total susceptibility function.

If we apply a *y*-polarized LPTL with a selected frequency of 1.1 THz, the susceptibility of the FE2, SP, and FE1 configurations are 731.5, 168.3, and 137.4, respectively. Under LPTL, the FE2 has lower Gibbs free energy than the FE1, indicating a FE1 to FE2 phase transition. In particular, when the laser intensity is increased to $2.2 \times 10^{10}$ W·cm$^{-2}$ (corresponding to electric field magnitude of 0.42 V·nm$^{-1}$), we find that $g_{FE_2} S_{FE_2} \leq g_{SP} S_{SP} < g_{FE_1} S_{FE_1}$. Thus, it is a barrier-free phase transition kinetics. As stated in the main text, such barrier-free phase transition could occur ultrafast and does not require latent heat. The process can avoid the conventional nucleation-and-growth kinetics, which is important for fast memory read/write technology.

## VIII. Supplementary Note 8: Direction dependent SHG susceptibility.

The space group of both GeSe and SnTe has $\mathcal{M}_x$ mirror symmetry. Thus, if the light polarization direction is along $\hat{\mathbf{n}} = (\cos\theta, \sin\theta)$, then the SHG susceptibility parallel (∥) and perpendicular (⊥) to $\hat{\mathbf{n}}$ can be written as (note that $\chi_{xxy} = \chi_{xyx}$)

$$\chi_\parallel(\theta) = (\chi_{yxx} + \chi_{xxy})\sin\theta\cos^2\theta + \chi_{yyy}\sin^3\theta \qquad (13a)$$

$$\chi_\perp(\theta) = (\chi_{yyy} - \chi_{xxy})\cos\theta\sin^2\theta + \chi_{yxx}\cos^3\theta \qquad (13b)$$

The rotation of orientation variant corresponds to change of polarization direction $\theta$, which indicates a ferroic order dependence of SHG susceptibility. In the main text, we plot $\chi_{yxx}(-2\omega;\omega,\omega)$, the $\chi_{yyy}$ and $\chi_{xxy}$ of both FE0-$\beta$-GeSe and FE1-$\alpha$-SnTe monolayers are plotted in Supplementary Fig. 9.

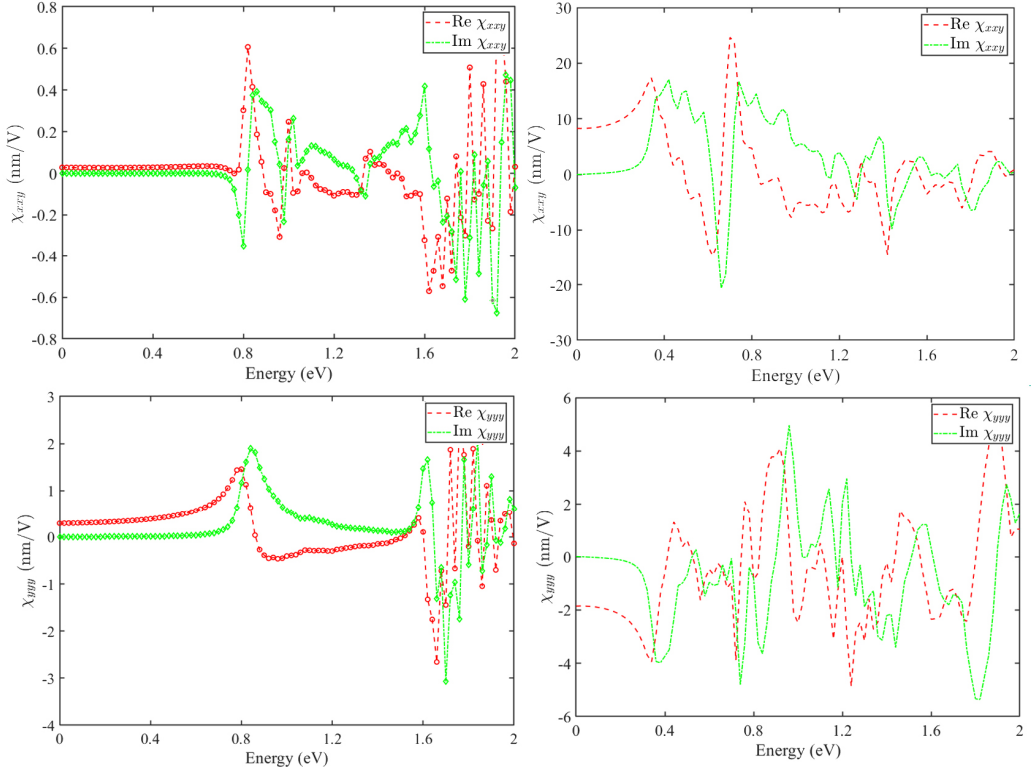

**Supplementary Figure 9.** Real and imaginary parts of SHG susceptibility of (left panel) FE0-$\beta$-GeSe and (right panel) FE1-$\alpha$-SnTe.

## IX. Supplementary Note 9: Microscopic mechanism of optics phonon interaction.

In the main text, the interaction between light and matter is discussed according to macroscopic thermodynamic approach, which could directly and straightforwardly provide the kinetics of phase transition. In this section, we briefly discuss its microscopic mechanism according to quantum mechanical perturbation theory[6]. The electric field can be written as

$$\mathbf{E}(t) = \mathbf{E}e^{i\omega t} \tag{14}$$

The semi-classical expression of induced electric moment between states $|l\rangle$ and $|m\rangle$ can be calculated as

$$\langle \mathbf{M} \rangle^{lm} = \int (\Psi_l^* \mathbf{M} \Psi_m + \Psi_m^* \mathbf{M} \Psi_l) \, d^3\mathbf{r} \tag{15}$$

The interaction between electric field and moment serves as an additional perturbation term in the Hamiltonian

$$H = H_0 - \mathbf{M} \cdot \mathbf{E} \tag{16}$$

The wavefunction $\Psi_l$ can be expanded according to perturbation theory,

$$\Psi_l = e^{-i\omega_l t} \left[ \psi_l + \frac{1}{\hbar} \sum_{\beta, r} \frac{\langle r | M_\beta | l \rangle}{\omega_r - \omega_l + \omega} \psi_r E_\beta e^{i\omega t} \right] \tag{17}$$

where $\hbar\omega_l$ and $\psi_l$ are the $l$-th eigenvalue and wavefunction of unperturbed system, respectively. $|r\rangle$ is a virtual state, $\beta = 1,2,3$ represents three Cartesian directions. The electric moment then takes the form

$$\langle \mathbf{M} \rangle_\alpha^{lm} = \sum_\beta \Pi_{\alpha\beta}^{lm}(\omega) E_\beta e^{i(\omega + \omega_l - \omega_m)t} \tag{18}$$

The $\Pi_{\alpha\beta}^{lm}(\omega)$ is transition polarizability, which indicates a non-coherent transition from $|l\rangle$ and $|m\rangle$

$$\Pi_{\alpha\beta}^{lm}(\omega) = \frac{1}{\hbar} \sum_r \left( \frac{\langle l | M_\alpha | r \rangle \langle r | M_\beta | m \rangle}{\omega_r - \omega_m + \omega} + \frac{\langle l | M_\beta | r \rangle \langle r | M_\alpha | m \rangle}{\omega_r - \omega_l - \omega} \right) \tag{19}$$

This process is illustrated in Supplementary Fig. 10.

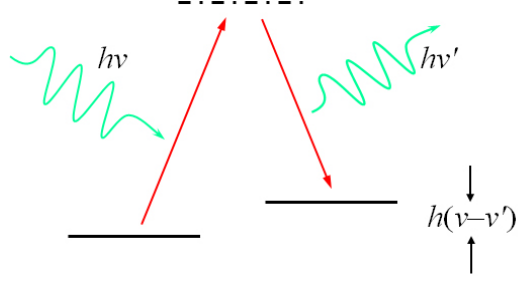

**Supplementary Figure 10.** Brief illustration of photon induced deformation. The solid lines represent real state that system could stay for sufficiently long time, unless there is dissipation. The dash-dotted line represents a virtual state. Classically, it corresponds to an oscillating ion or electron subsystems.

According to Placzek adiabatic approximation, the total wavefunction $\Psi$ can be divided into ion wavefuntion $\varphi$ and electron wavefunction $\phi$,

$$\Psi_{n\mu}(x,X) = \varphi_{n\mu}(X)\phi_n(x,X) \tag{20}$$

Here $n$ and $\mu$ are electronic and ionic quantum numbers, $x$ and $X$ represent electron and ion positions, respectively. Thus, the $|l\rangle$ and $|m\rangle$ states can be written as $|n\mu\rangle$ and $|n'\mu'\rangle$, respectively. The electronic ground state is $n=0$. The transition polarizability can be written as

$$\Pi_{\alpha\beta}^{\mu\mu'}(\omega) = \frac{1}{\hbar}\sum_{n'',\mu''}\left(\frac{\langle 0\mu|M_\alpha|n''\mu''\rangle\langle n''\mu''|M_\beta|0\mu'\rangle}{\omega_{n''\mu''}-\omega_{0\mu'}+\omega} + \frac{\langle 0\mu|M_\beta|n''\mu''\rangle\langle n''\mu''|M_\alpha|0\mu'\rangle}{\omega_{n''\mu''}-\omega_{0\mu}-\omega}\right) \tag{21}$$

The contribution can be divided into two parts, namely, electronic and ionic contributed transition polarizability

$$\Pi_{\alpha\beta}^{\mu\mu'}(\omega) = \Pi_{\alpha\beta}^{\mu\mu',(\text{el})}(\omega) + \Pi_{\alpha\beta}^{\mu\mu',(\text{ion})}(\omega) \tag{22}$$

For the ionic contributed part, the electron always stays at its ground state when the ion transits from $|\mu\rangle$ to virtual $|\mu''\rangle$ and drops back to $|\mu'\rangle$. Thus,

$$\Pi_{\alpha\beta}^{\mu\mu',(\text{ion})}(\omega) = \frac{1}{\hbar}\sum_{\mu''}\left(\frac{\langle\varphi_{0\mu}|M_\alpha|\varphi_{0\mu''}\rangle\langle\varphi_{0\mu''}|M_\beta|\varphi_{0\mu'}\rangle}{\omega_{0\mu''}-\omega_{0\mu'}+\omega} + \frac{\langle\varphi_{0\mu}|M_\beta|\varphi_{0\mu''}\rangle\langle\varphi_{0\mu''}|M_\alpha|\varphi_{0\mu'}\rangle}{\omega_{0\mu''}-\omega_{0\mu}-\omega}\right)$$

$$\tag{23}$$

As for the electron contributed part, usually the incident optical frequency lies above the phonon frequency range, thus the ion virtual state can be omitted, and $\omega_{n''\mu''} - \omega_{n\mu} \cong \omega_{n''} - \omega_n$. It thus takes the form of

$$\Pi_{\alpha\beta}^{\mu\mu',(\text{el})}(\omega) = \frac{1}{\hbar}\sum_{n''}\left(\frac{\langle\varphi_{0\mu}\phi_0|M_\alpha|\phi_{n''}\rangle\langle\phi_{n''}|M_\beta|\phi_0\varphi_{0\mu'}\rangle}{\omega_{n''}-\omega_0+\omega} + \frac{\langle\varphi_{0\mu}\phi_0|M_\beta|\phi_{n''}\rangle\langle\phi_{n''}|M_\alpha|\phi_0\varphi_{0\mu'}\rangle}{\omega_{n''}-\omega_0-\omega}\right).$$

(24)

When the light frequency is above the phonon frequency, $\Pi_{\alpha\beta}^{\mu\mu',(\text{ion})}(\omega) \ll \Pi_{\alpha\beta}^{\mu\mu',(\text{el})}(\omega)$, so that the ion contributed transition polarizability is negligible (also $\mu \cong \mu'$). If the light frequency lies in the phonon frequency range (~ THz), the contributions from electron and ion are similar, so that they should be considered simultaneously. Note that in this case, since the electronic bandgap is larger, the $\Pi_{\alpha\beta}^{\mu\mu',(\text{el})}(\omega)$ is almost a constant, $\Pi_{\alpha\beta}^{\mu\mu',(\text{el})}(\omega) = \Pi_{\alpha\beta}^{\mu\mu',(\text{el})}(\omega = 0)$. This process has no light absorption (up to a real electron or phonon state), thus there is no significant waste energy.